# Boiling transitions during droplet contact on superheated nano/micro-structured surfaces

**Navid Saneie*[1], Varun Kulkarni*[1], Kamel Fezzaa[2], Neelesh Patankar[3], Sushant Anand‡[1]**

[1]Mechanical & Industrial Engineering, University of Illinois at Chicago, Chicago, IL 60607
[2]Advanced Photon Source, Argonne National Laboratory, Argonne, IL 60439
[3]Mechanical Engineering, Northwestern University, Evanston, IL 60208
*Equal contribution
‡ Corresponding author's e-mail address: sushant@uic.edu

**Manipulating surface topography is one of the most promising strategies for increasing the efficiency of numerous industrial processes involving droplet contact with superheated surfaces. In such scenarios, the droplets may immediately boil upon contact, splash and boil, or could levitate on their own vapor in the Leidenfrost state. In this work we report the outcomes of water droplets coming in gentle contact with designed nano/micro-textured surfaces at a wide range of temperatures as observed using high-speed optical and X-ray imaging. We report a paradoxical increase in the Leidenfrost temperature ($T_{LFP}$) as the texture spacing is reduced below a critical value (~10 μm). Although droplets on such textured solids appear to boil upon contact, our studies suggest that their behavior is dominated by hydrodynamic instabilities implying that the increase in $T_{LFP}$ may not necessarily lead to enhanced heat transfer. On such surfaces, the droplets display a new regime characterized by splashing accompanied by a vapor jet penetrating through the droplets before they transition to the Leidenfrost state. We provide a comprehensive map of boiling behavior of droplets over a wide range of texture spacings that may have significant implications towards applications such as electronics cooling, spray cooling, nuclear reactor safety and containment of fire calamities.**

Droplet interactions with superheated surfaces are ubiquitous, playing a decisive role in some of the most critical applications such as spray-cooling, electronics cooling, combustion processes in automobile engines, fire extinguishing, and metal quenching[1]. Such interactions share many attributes with pool boiling on superheated surfaces while also having some unique features that distinguish them apart. For example, at low superheat temperatures, rapid bubble formation occurs in spreading droplets analogous to nucleate pool boiling. Similarly to pool boiling, with an increase in the surface superheat, hydrodynamic instabilities play an increasingly important role in droplet-solid interactions, but unlike pool boiling, such instabilities manifest themselves by disintegrating the droplet in what is typically referred to as "splashing regime"[2,3]. Finally, beyond a critical temperature referred to as the Leidenfrost temperature (hereafter denoted by $T_{LFP}$), the droplet no longer remains in direct contact with the heated surface but is supported by an intervening vapor layer – a condition analogous to film boiling in bulk liquids[4-6]. Researchers have shown that droplets in this state can be used for nanoparticle synthesis[7] and can self-propel unidirectionally by manipulating the surface structure[8,9]. However, for many of the cooling related applications mentioned previously, the long-standing goal has been to avoid/delay the catastrophic consequences of inefficient heat transfer in the Leidenfrost regime and expand the nucleate/splashing regime to take advantage of the high heat fluxes associated with them[10-12].

Because of such widespread applicability, there is tremendous interest in understanding the different boiling regimes. This however is a formidable challenge considering the complexity of interactions and the large number of factors governing them. For example, the $T_{LFP}$ is dependent upon the chemical and thermophysical[13,14] properties of the droplets, the boiling/impact conditions, and topography of the solid surface. Textures with micro/nano or hierarchical features can promote drag reduction[15-17] if they are hydrophobic by expediting the occurrence of $T_{LFP}$, whereas extremely rough hydrophilic surfaces can aid efficient cooling[18,19] by delaying the $T_{LFP}$ due to high wettability and increased number of nucleation sites[20]. Even for the same substrate and impacting liquid, the $T_{LFP}$ is a dynamic quantity, varying with and depending upon impacting conditions[4,21,22]. Although quantitative predictions of the dynamic $T_{LFP}$ of impacting droplets on smooth surfaces are yet to be established, at least for the case of small impacting velocities, it has been shown that parameters such as the vapor layer thickness below the droplet and its pressure can be estimated by balancing the forces acting on the droplet[5,6]. This force balance approach also appears to hold for micro-rough textured surfaces to predict $T_{LFP}$[23,24]. However, it is unclear whether such an approach can be extended to nanoscale features at particular texture spacing. Studies on droplets impacting nanowires suggest an increase in dynamic $T_{LFP}$[25] with the observance of vigorous explosions at the solid-liquid interface[26-29] before reaching the Leidenfrost state, indicating a localized pressure build-up below the droplets. It remains unclear whether such features are unique to such surfaces, and what role (if any) they play in the determination of $T_{LFP}$. Beyond the Leidenfrost state, while some studies have investigated other regimes during droplet impact[21,22,30,31], a fundamental understanding of how the surface texture affects the transitions between different regimes remains elusive.

In this study, we have performed a systematic investigation of boiling behavior on textured surfaces with spacings from 1 μm to 100 μm to understand the mechanisms leading to the various boiling morphologies observed at various temperatures and spacings. We especially seek answers to several crucial questions such as - how the boiling behavior is altered when surface roughness is high; whether the Leidenfrost temperature asymptotes to a constant value at these lower spacings; are new boiling morphologies observed prior to the Leidenfrost state and what are the governing mechanisms defining various boiling transitions at these texture spacings. To answer these, we performed gentle drop impact experiments and visualized their behavior using high-speed optical imaging. We also conducted the experiments at Argonne National Lab[32] visualizing the drop impact behavior using high-speed X-ray phase contrast imaging (XRPCI) that has previously been used in studying variety of fluid phenomena[33], and recently also in studying of boiling[34-36]. By virtue of the two visualizing techniques, we unearthed a new boiling regime unbeknownst until now as the surface texture spacing is reduced below a critical value (~10 μm in the present study). This newly discovered regime has profound implications towards our understanding of boiling as it indicates that hydrodynamic instabilities may interfere with achieving better cooling and also negate the increase in surface hydrophilicity effects as the surface texture spacings are reduced.

## Results

**Leidenfrost temperature on hydrophilic nano/micro-structures surfaces.** Droplet behavior on hot textured surfaces in ambient environmental conditions primarily depends upon the texture (post) spacing, the impact velocity ($V_d$), and the droplet diameter ($D_d$). The effects of the latter two factors can be captured using the droplet Weber number, the ratio of inertial and surface tension forces, given by $We_d = \rho_l V_d^2 D_d / \sigma_l$ where $\rho_l$ and $\sigma_l$ correspond to the liquid density and surface tension respectively. We first used high-speed optical imaging to find the dynamic $T_{LFP}$ on fabricated silicon micropillars, 10 μm both in diameter (*a*) and height (*h*) with spacings between the posts (*b*) as 1.5, 2.5, 5, 10, 50, 75 and 100 μm. Due to fabrication limitations for samples with large diameter to spacing ratio, Nano-grass (NG) samples with an average spacing of approximately 1 μm were also fabricated to study the effect of smaller micropillar spacings (Fig. 1a). The height of the NG samples was measured to be ~10μm (Supplementary Figure S1). The textured samples were rendered hydrophilic by plasma treatment before performing the experiments. While previous works have studied the variation of $T_{LFP}$ during droplet impact[3,30,31], we concentrated on low $We_d$ impacts (ranging from 1 to ~8) by varying the droplet sizes and the impact height ($H_o$) ranging from ~0 to 5 mm above each textured substrate. Textured samples were placed on a heater, and the temperature was gradually increased with 5 °C degrees increments, before each experiment. Upon reaching steady-state conditions, the behavior of droplets upon contact was recorded using a high-speed camera (Phantom AX100) at 10000 fps.

In pool boiling experiments, the onset of stable film boiling regime is usually defined as LFP, representing the minimum heat flux in the boiling curve. In the case of liquid droplets/puddles deposited on a heated surface, "no local contact" or "maximum evaporation time" are among common perspectives found in literature to present the Leidenfrost temperature. Another perspective to look at the LFP for the case of liquid drops coming to contact with heated surfaces, specially in lower We number regime, is the bouncing behavior of the deposited drops. This may be perceived as a sign of no local contact and a stable vapor film between the liquid and substrate. Getting help from high speed imaging techniques (as high as ~20000fps in our case), we can analyze the details of bouncing behavior to guarantee that there was no contact and no further sign of boiling at the temperature we measured for Leidenfrost temperature. Therefore, the smallest substrate temperature at which bouncing was observed upon first droplet contact without splashing was noted as $T_{LFP}$. Further details on the experimental protocol and setup are provided in the methods section and Supplementary Fig. S1a.

Fig. 1b presents the variation of dynamic $T_{LFP}$ as a function of micropost spacing from 1 μm (nanograss) to 100 μm for different $We_d$. The first noteworthy observation is that for a given texture morphology, even a small increase in $We_d$ led to a significant delay in the Leidenfrost state. While previous studies have observed this behavior during high $We_d$ impact[21,30] and only on surfaces with larger micropost spacings (*b*>10 μm), our results indicate that this dependence extends to texture spacings as small as 1μm. With regards to the variation of $T_{LFP}$ with post spacing, we found that with increasing the spacings above ~10 μm, our experimental results agree with previously established increasing trend of $T_{LFP}$ for gentle deposition on microstructures[20,23]. However, intriguingly, we found that decreasing post spacings below ~10 μm, $T_{LFP}$ increased again with the minima for the $T_{LFP}$ occurring at a spacing of approximately 10 μm, irrespective of the droplet impact $We_d$. This phenomenon suggests a crucial connection between $T_{LFP}$ on microstructures expanding upon limited previous work that reported high Leidenfrost temperatures for rough surfaces and porous materials[37,38].

**Transition between different boiling regimes.** To understand the reversal of $T_{LFP}$, we examined the nature of droplet boiling dynamics just before the LFP on each micro-structured surface as captured by high-speed optical imaging. We focused our attention to gentle deposition experiments ($We_d$ ~1) to eliminate the dynamically triggered behavior of impacted drops. Analyzing the experimental videos, it became evident that the nature of boiling dynamics of drops was distinct for different spacings when the surface temperature was just below the $T_{LFP}$ (Fig. 2a, also see Supplementary Video 1). On surfaces with *b*>10 μm, the droplet spreading was accompanied by rapid boiling and ejection of micro-droplets along the apparent droplet contact line, which has been previously identified as splashing regime. Similar behavior was also observed on surfaces with *b*<10 μm; however, on such surfaces, a vapor cloud rising rapidly within the droplet formed simultaneously with the splashing around the contact area.

The challenges inherently associated with optical visualization limited thorough identification of critical details of the boiling behavior inside the droplets. To overcome such challenges, we visualized gentle drop impact using high-speed XRPCI[32] (see the methods section and Supplementary Fig. S1c for the details of the experimental setup). We studied the boiling behavior of gently deposited droplets ($We_d$~1) over three surfaces with spacings *b* = 1 (NG), 10 and 100 μm, and temperatures above the saturation temperature ($T_{sat}$ =100 °C) of water. On each sample, the surface temperature was slowly increased from $T_{sat}$ until $T_{LFP}$ was reached and droplet contact for each temperature was visualized using high-speed XRPCI. Based on the visual observations, we were able to conclusively identify four distinct droplet boiling behaviors on surfaces as a function of surface temperature and texture. As surface temperature ($T_s$) exceeded few degrees above $T_{sat}$, the nucleate boiling regime characterized by bubble nucleation and rapid bubble growth was observed on all the surfaces (see images marked ■ in Fig. 2b, also see Supplementary Video 2). As $T_s$ was increased, the splashing regime characterized by vigorous ejection of droplets from the droplet-substrate contact line was observed on all the three surfaces (see images marked ★ in Fig. 2b, also see Supplementary Video 2). While for droplets impacting at high $We_d$ the splashing is mainly ascribed to the drop impact inertia[29,30], for our experiments the transition to the splashing boiling regime occurs due to the instabilities induced by the vapor partially escaping through the micropillar texture. For smaller spacings ($b < 10$μm), further increasing the temperature beyond the splashing regime led to a remarkable, abrupt vapor explosion giving rise to a mushroom-like jet, reminiscent of a vortex ring, inside the drop (see images marked ▲ in Fig. 2b, also see Supplementary Video 2). Henceforth, this behavior is referred to as the jet-splashing regime. This explosion is the result of the accumulated pressure below the droplet, overpowering the downward pressure of the drop comprised of the weight of the drop and the capillary force resulting from the presence of the microstructure, the detailed dynamics of which will be discussed later. It is instructive to point out that the bubbles formed after the explosion are the remnants of the air (vapor) trapped inside the droplet and are not formed solely as a consequence of the nucleate boiling. Further increasing the surface temperature beyond the $T_{LFP}$ led to the Leidenfrost state as discussed before (marked ● in Fig. 2b and Supplementary Video 2). In this state we could also visualize the evolution of the "dimple" below the droplet coming in contact with the microstructured surfaces. This observed dimple is indicative of the pressure build-up below the evaporating droplet.

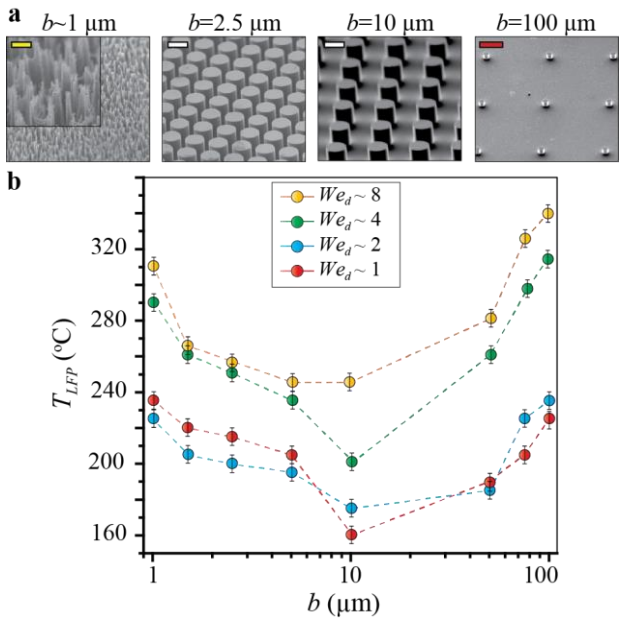

**Figure 1| Leidenfrost temperature as a function of micropillar spacings and Weber number.** (**a**) SEM images of the microstructured surfaces (NG, b=2.5, 10 and 100μm) studied in this work. Yellow, white and red scale bars are 1, 10 and 50μm, respectively (**b**) Variation of $T_{LFP}$ as a function of post spacing (b=1.5-100 μm and b~1μm for nano-grass samples) and the Weber number. Weber number is calculated based on the impact velocity of deposited droplets.

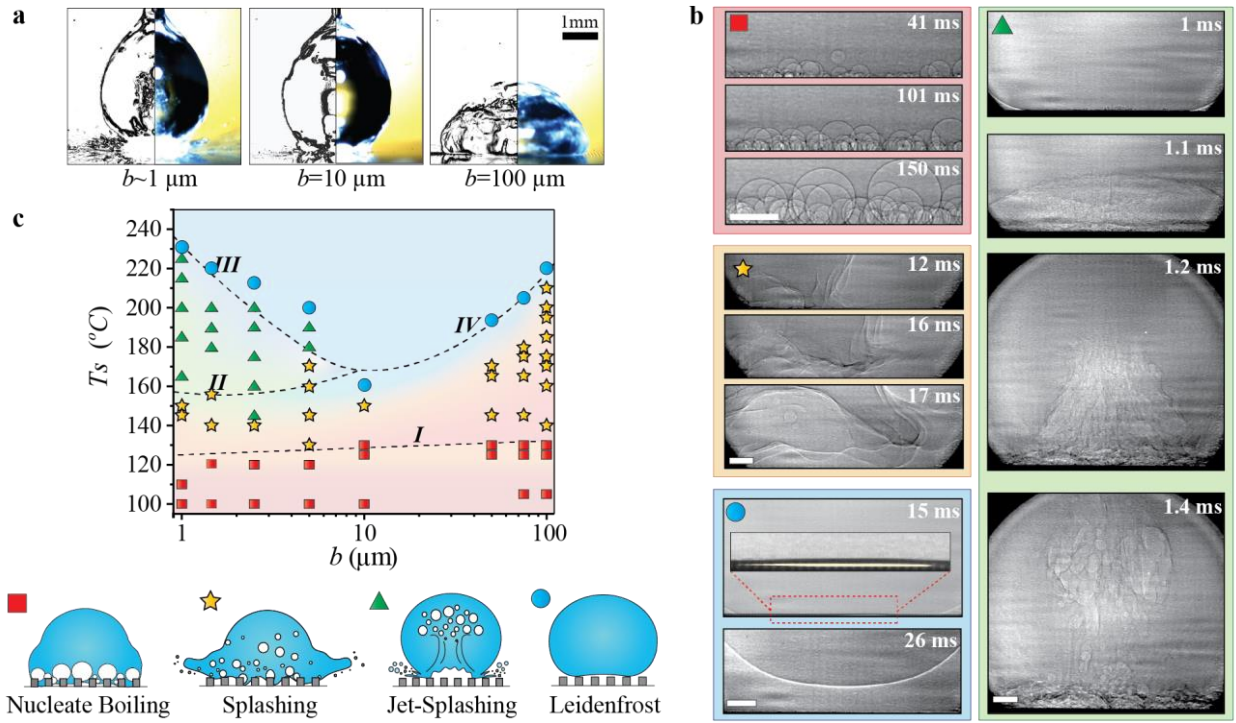

**Figure 2 | Boiling behavior of droplets contacting superheated surfaces.** (a) Optical images showing water droplet boiling behavior just prior to the Leidenfrost temperature on NG, 10 μm and 100 μm microstructure spacing samples. Images were enhanced in ImageJ to visualize boiling dynamics within droplets. (b) X-Ray images showing the four types of droplet phenomenon seen in our experiments. The nucleate (red) and splashing (yellow) boiling regimes are observed for all the spacings. The jet formation (green) was only observed on the samples with micropillars spacing less than 10μm. A dimple underneath the droplet at Leidenfrost regime (blue) is formed, before any bouncing behavior, due to the pressure accumulation. The scale bar for X-Ray images is 200μm. (c) The regime map for the gentle deposition experiment ($We$~1) with the dashed lines showing qualitative transitions curves distinguishing between all the boiling regimes.

While violent boiling of droplets and liquid jets on nanowire surfaces has been observed[27-29], the true nature and the extent of such a regime as a function of texture properties have not been investigated previously. Our observations engender further questions such as, whether the transitions between different stages of boiling (for example, from nucleate to splashing) vary as a function of the surface temperature and the micro-texture spacing. Thus, we sought to obtain a comprehensive map of all the transition regimes on superhydrophilic textured surfaces by focusing exclusively on identifying the key signatures of four boiling behavior identified above, observed immediately after the droplet comes to contact with the textured substrate. The results of this study are shown in Fig. 2c and the images showing the underlying characteristic behavior appear in Supplementary Fig. S2. In the ensuing discussion, we have named the transition between nucleate & splashing, splashing & jet-splashing, jet-splashing & LFP, and splashing & LFP as Curve I, II, III, and IV respectively. It should be noted that one regime gradually transitions to the other, and hence these boundaries are not sharp but spread over a small temperature range. The first notable revelation of the regime map (Fig. 2c), is that the transition from nucleate to splashing regime (Curve I) is nearly independent of the texture spacing. As we will discuss later in detail, this behavior stems from the dominant role played by the Kelvin-Helmholtz instabilities in the droplets, induced by the vapor shear-flow. Furthermore, this map reveals the boundaries for the so-called "jet formation regime" (Curve II and III), which was distinctive at smaller micropillars spacings ($b<10$μm). Observing these boundaries, we also propose that the gradual rise in the $T_{LFP}$ as we move towards such small spacings is due to particular mechanisms which are discussed in the remainder of the text.

**Boiling versus splashing time-scales.** Below the $T_{LFP}$, the incipience of boiling, splashing and jetting (for surfaces with $b<10$ μm) visually appears to be almost simultaneous. Nonetheless, we found that from the moment the droplet comes into contact with the surface ($t=0$), there is a discernible difference in times at which these phenomena are observed. Using the optical imaging, we obtained the estimated times for splashing ($\tau_{s,exp}$) and jetting ($\tau_{jet,exp}$). The boiling time ($\tau_b$) i.e., the time for bubbles to become noticeable within the droplet upon its contact with the surface was obtained from high-speed X-ray imaging. The different boiling behaviors just prior to the $T_{LFP}$ are shown for two extreme cases ($b$~1 μm and 100 μm) in Fig. 3a using optical and X-Ray imaging. For smaller spacings ($b<10$μm), optical image analysis reveals that the droplets show splashing behavior before any jet formation for all the cases involving jet-splashing (Fig. 3a and also see Supplementary Fig. S3) suggesting that the splashing behavior is the dominating mechanism on such surfaces. X-ray imaging showed that in such cases, bubble nucleation and growth was inconspicuous (Fig. 3a). On the other hand, for spacings larger than 10 µm, X-ray imaging showed distinct bubble nucleation and growth (Fig. 3b).

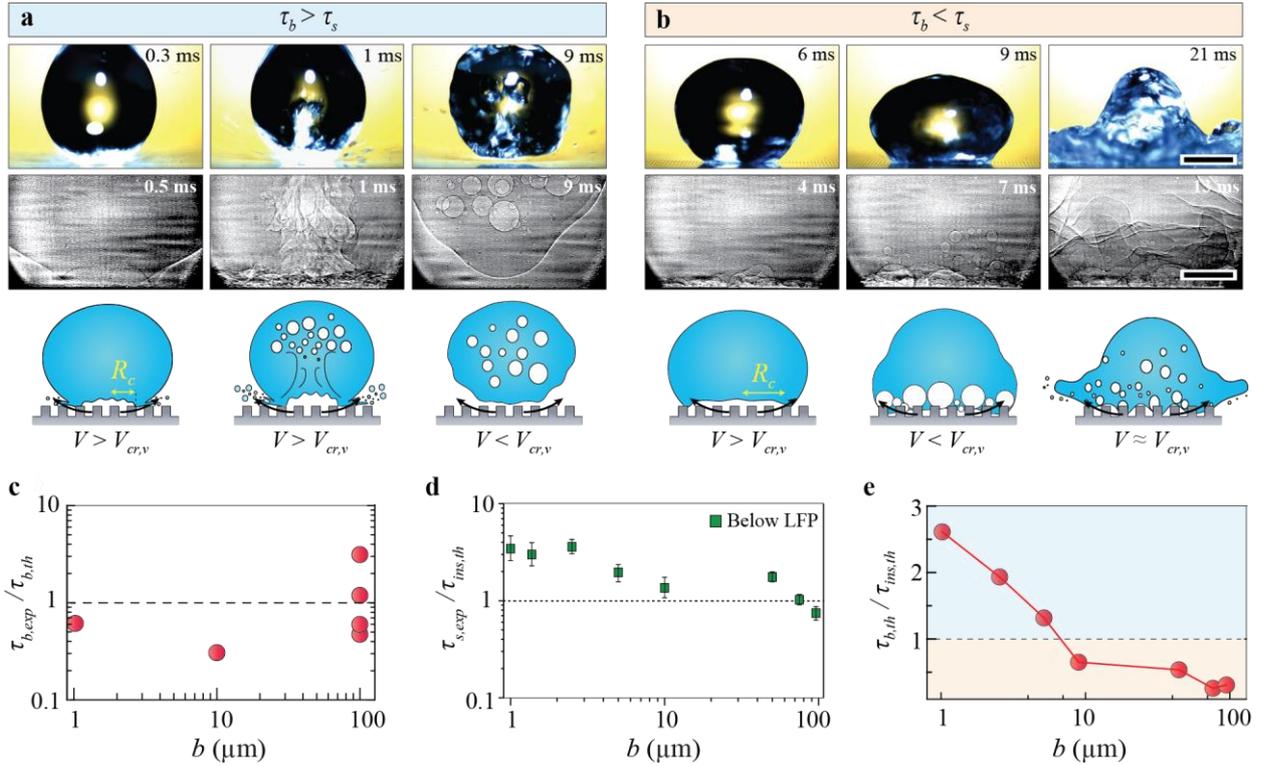

**Figure 3| Time-scale comparison between initiation of splashing and nucleate boiling/bubbling regimes**. (**a**) Optical and X-Ray Image sequence of boiling behavior on $b$=1 µm showing occurrence of splashing followed by jet formation. This behavior was observed for samples with $b$<10 µm. (**b**) Boiling behavior on $b$=100 µm showing occurrence of bubble nucleation and growth followed by splashing behavior. The scale bar for optical images and X-Ray images are 1mm and 0.5 mm respectively.(**c, d**) Comparison between theoretical and experimental time-scales of boiling and splashing behaviors. Error bars are calculated based on the imaging speed and resolution. (**e**) Comparison between theoretically predicted time-scales of boiling and splashing initiation as a function of micropillar spacing.

To rationalize the above behavior, we predicted the boiling time-scale ($\tau_{b,th}$) for the impacting droplets. The temperature distribution within a droplet coming into contact with a solid surface held at $T_{s,i}$ can be obtained using the 1D transient energy equation (Supplementary Note 1) as given by:

$$T(x,t) = T_{c,i} + (T_{l,i} - T_{c,i}) \, \text{erf}\left(\frac{x}{2\sqrt{\alpha_l t}}\right) \quad (1)$$

where $T_{l,i}$ is the initial droplet temperature, $x$ is the distance from the solid surface, $T_{c,i}$ is the contact interface temperature and $\alpha_l$ is the diffusivity of water. The contact interface temperature ($T_{c,i}$)[39] can be expressed as:

$$T_{c,i} = \frac{T_{l,i}\sqrt{k_l \rho_l c_l} + T_{s,i}\sqrt{k_s \rho_s c_s}}{\sqrt{k_l \rho_l c_l} + \sqrt{k_s \rho_s c_s}} \quad (2)$$

where $k$ is the thermal conductivity and $c$ is the specific heat capacity of the medium. Subscripts $s$ and $l$ represent solid and liquid respectively and $i$ refers to the initial temperature before the contact. Following the nucleation theory, the nucleation and thereafter growth of bubbles within the droplet occurs when the temperature at the thermal layer thickness ($\delta_{th}$) reaches the saturation boiling temperature of the liquid, $T_{sat}$[40,41]. Considering the nucleation between and on top of the microposts (see Supplementary Fig. N1), we estimate $\delta_{th}$ to be ~10 µm. It is worth mentioning that this distance does not considerably affect the boiling time scale if taken within the same order of magnitude. Setting $x=\delta_{th}$ and $T(\delta_{th}, \tau_b) =$ $T_{sat}$ in Eqn. (1) and using Eqn. (2), the boiling time-scale can be expressed as:

$$\tau_{b,th} = \frac{\delta_{th}^2}{4\alpha_l \psi}; \qquad \psi = \left(\text{erf}^{-1}\left(\frac{T_{sat} - T_{c,i}}{T_{l,i} - T_{c,i}}\right)\right)^2 \quad (3)$$

A comparison between the theoretical and experimentally determined (using X-ray imaging) time-scales for boiling inception for different temperatures and spacings as given in Fig. 3c shows that they are considerably well-matched. Note that for the cases with surface temperatures much larger than the $T_{sat}$, the matching accuracy would increase upon using the effective thermal conductivity $k_{eff}$ of the solid substrate (see Supplementary Note. 2) to account for the droplet being in contact with its vapor and the solid surface.

To understand the mechanisms leading to splashing and predict the time-scale at which this outcome is observed ($\tau_{s,exp}$), the events preceding it need to be carefully analyzed. X-Ray recordings showed that before the splashing behavior, the droplet surface displayed disturbances manifesting in the form of waves (Fig. 2b★ and Supplementary Video 3). These interfacial waves are driven by the shear forces arising from the vapor flow underneath the evaporating drop and are stabilized by the liquid surface tension and gravity - a competition demonstrating the Kelvin-Helmholtz instabilities[42,43]. At low surface superheats, the vapor flow velocity is greatly decreased due to reduced evaporation and therefore is insufficient to perturb the interface and cause splashing. As temperature increases, eventually a critical condition is reached where the surface tension forces can no longer contain the instabilities, and they ultimately grow, leading to thinning of the droplet at its periphery. The eventual fragmentation of the thinned periphery of the liquid droplet then occurs due to the surface tension driven Rayleigh-Plateau (RP) instability[44,45]. Thus, the splashing time of the droplet encompasses the time-scale for manifestation of the KH (interfacial) instabilities ($\tau_{ins,th}$) and the RP initiation time-scale ($\tau_{RP}$). However, the initiation time-scale is much larger than the corresponding time-scale for RP instability ($\tau_{ins,th} >> \tau_{RP}$, see Supplementary Note 3) which implies that once the critical instability waves appear, micro-droplets are generated at the circumference of the contact area instantaneously ($\tau_{s,exp} \approx \tau_{ins,th}$)[46]. It can be shown that the KH instability will grow on the droplet only beyond a critical velocity ($V_{cr,v}$) of the vapor ($v$) flow given by:

$$V_{cr,v} > \left(\frac{4\sigma_l \rho_l g}{\rho_v^2}\right)^{1/4} \quad (4)$$

Inserting values for the liquid and vapor properties from the experimental conditions, the average value for this velocity is approximately 8 m/s [47](see Supplementary Note. 3). Experimentally, the vapor velocity ($V_v$) can be estimated using the ejected microdroplets as tracers[48] (see Supplementary Fig. N5). These measurements confirmed that splashing behavior is only observed once the vapor velocity exceeds the critical value ($V_v > V_{cr,v}$). It is worth mentioning that the ejected droplets in the nucleate boiling regime is indeed due to the bubble bursting[6,29] and their velocity is much lower than the critical vapor velocity for KH instability. However, in the jet-splashing and splashing regime, the sudden explosion and KH instabilities drive the ejected droplets with a velocity greater than $V_{cr,v}$. The vapor velocity $V_v$ is indeed a function of post spacing. Observing the liquid-solid interface using the X-Ray imaging (See supplementary video 4), we clearly see that the liquid surface is never uniformly seated on nor completely imbibed in the posts' gaps. Therefore, using theoretical values of vapor velocity calculated using force balance approach, where the droplet is assumed to uniformly sit on top of the posts can be very misleading.

For the vapor shear force-induced interfacial waves, the velocity of wave propagation $v_P$ can be defined as ($\omega/\Gamma$) where $\omega$ is the growth rate of the wave[43,49]. This ratio can also be expressed as ($\omega/\Gamma$)$^2$ = $B$-$\Gamma A$ where $\Gamma$ is the wave number, $A = \sigma_l/\rho_l$ and $B = (\rho_v/\rho_l)V_v^2$. Differentiating the wave equation and setting it equal to zero, we can find the maximum growth rate $\omega_{max}$. Recognizing that $\tau = \omega^{-1}$, we arrive at the expression below for KH instability time-scale at which the instabilities cause the splashing of micro-droplets:

$$\tau_{ins,th} \approx \sqrt{\frac{27A^2}{4B^3}} \quad (5)$$

The time-scale for KH instability obtained from theory matches well with experimental data (Fig. 3d).

Having obtained the boiling and splashing time-scales, we then used these relations and the experimental $T_{LFP}$ and $R_c$, the contact area radius, to estimate the dominating mechanism for different spacings just prior to the $T_{LFP}$. This analysis shown in Fig. 3e indicates that for smaller spacings ($b < 10$ µm) the dynamics are primarily governed by KH instabilities rather than boiling – just as observed in experiments (Fig. 3a). For larger spacings ($b > 10$ µm), the vapor speed never exceeded the critical value before the initiation of boiling at the interface, and hence on such surfaces the dynamics were dictated by the bubble nucleation as found in our experiments (Fig. 3b). These comparisons are not only true for temperatures below LFP at smaller spacings, but also for all of the temperatures above the nucleate boiling regime and are critical in understanding the transitions between different regimes as explained in the next sections. It should also be noted that the time-scale for vapor convection cooling in the gentle droplet impact is significantly larger than the time-scale (see Supplementary Note 4) for instabilities, implying that the increase in LFP at small texture spacings is not because of forced convection effects.

**From nucleate boiling to splashing regime.** To obtain the temperature at which the drop transitions from the nucleate boiling regime to the splashing regime, we consider the balance between the evaporation rate from the droplet base ($\dot{m}_{gen}$) and the escape rate ($\dot{m}_{esc}$) from within the pillars. The vapor mass $\dot{m}_{gen}$ generated can be evaluated from a steady state energy balance which amounts to equating the heat conduction to the latent heat for vaporization, and is given by:

$$\dot{m}_{gen} = \frac{k_{eff}\pi R_c^2 \Delta T}{z L} \quad (6)$$

where $\Delta T$ is the substrate superheat, $R_c$ is the contact area radius at the time of splashing, $L$ is the latent heat of vaporization for water, $z$ is the thickness of the vapor film in the splashing regime which is assumed to be equal to the height of the micropillars ($h = 10$µm) since the droplet in nucleate/splashing regime is primarily in contact with the textured surface. The droplet contact radius ($R_c$) at the instant of the droplet showing boiling/splashing varies as a function of spacings and temperatures (see Supplementary Fig. S4). As the generated vapor escapes from within the texture, its escape rate is given as $\dot{m}_{esc} = 2\pi h \rho_v V_{cr,v} R_c(b,T)$ where $V_{cr,v}$ was chosen as velocity scale, since the splashing initiates once the vapor velocity exceeds its value. Equating the mass loss due to evaporation ($\dot{m}_{gen}$) with the vapor escape rate ($\dot{m}_{esc}$) from the edge of the contact line, the minimum temperature needed for the transition from the nucleate boiling regime to splashing regime is estimated as

$$T_{c,I} \approx T_{sat} + \frac{2\rho_v h^2 V_{cr,v} L}{k_{eff}(b) R_c(b,T)} \quad (7)$$

The predictions from Eqn. (7) were found to be in an excellent match with the experimentally observed temperatures at which the boiling/splashing transition was observed (Fig. 4).

**From splashing to jet-splashing regime.** For the case of smaller spacings ($b<10$ µm), a gradual increase in temperature in the splashing regime is accompanied by vapor pressure build up underneath the drop due to the rapid increase in evaporation and decrease in permeability of the textured surface[50,51]. Hypothetically, the resulting partial blockage of vapor flow due to droplet imbibition (see Supplementary Video 4) further aids this pressure accumulation, the consequences of which can be seen in form of a dimple underneath the droplet. As one would anticipate, this vapor pressure increase cannot possibly continue unabated. Eventually, the droplet can no longer sustain this pressure, leading to a rapid deformation of the liquid-air drop interface which culminates in a mushroom-like jet formation inside the drop (Fig. 2b). Such deformation bears a remarkable resemblance to liquid films, for our case the liquid drop, which is accelerated along their surface normal (in the out-of-plane direction). These films are unstable due to Rayleigh-Taylor (RT) instability since the lighter fluid (vapor) rapidly accelerates into a denser fluid (water) ultimately leading to puncturing of the liquid mass. Similar liquid rupturing has been reported in the context of drop atomization and impact where RT instability has successfully explained the observed flow features[52]. It is critical to differentiate this from the classical RT instability where the instability arises as a consequence of gravitational acceleration and is typically seen in the case of unfavorable density stratification[53]. The maximum amplified wavelength[54] ($\lambda_c$) corresponding to RT instability is given by:

$$\lambda_c = 2\pi \sqrt{\frac{3\sigma_l}{\rho_l a_{lv}}} \quad (8)$$

Here $\rho_l$ is density of the liquid, $\sigma_l$ is the surface tension and $a_{lv}$ is the acceleration of the vapor-liquid interface. For our case, we have a competition between the rapid acceleration destabilizing the interface and surface tension which tends to stabilize it[55]. Such rapid acceleration of the interface is due to the pressure acting below the droplet. It is worth recognizing that the acceleration of the liquid interface, outside of the jet-splashing regime, should be low enough to not lead to its rupture. Such low accelerations can arise in two scenarios (i) when the velocity is low due to small evaporation rates at smaller spacings (curve II, and described here) (ii) when the area ($\sim R_c^2$) over which the pressure force acts, and the thickness of the vapor cushion are large, implying lower pressure underneath the drop (and thereby diminished value of interfacial acceleration, curve III). The second scenario is discussed in the context of curve III in the subsequent section. In our explanation here, we restrict our attention to the conditions corresponding to (i) and curve II.

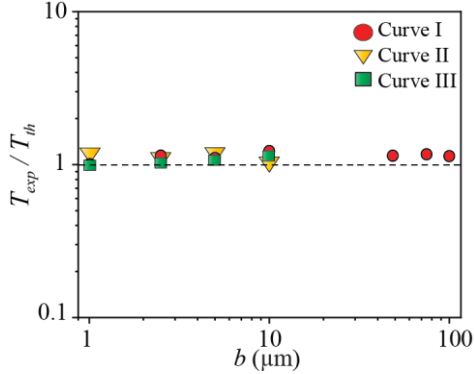

**Figure 4| Boiling transition curves with respect to the spacing between the micropillars.** The graph shows the transition from nucleate boiling to splashing regime (red) as well as the splashing to jet formation regime (yellow) and finally the transition from jet formation to Leidenfrost state (blue). The theoretical values agree with the experimentally measured temperatures for every transition

In view of the discussion above, we calculate the temperature required to transition from splashing to jet-splashing. However, we notice that the appearance of an unstable interface complicates any estimation of pressure which is required to make this calculation. Alternatively, we used the criterion for the waves to grow unabatedly which translates to a critical aerodynamic Weber number ($We_{cr} = \rho_v U_{cr}^2 d_o / \sigma_l$) ~10 to calculate the velocity of vapor flow required to cause such a rupture[48]. Employing this critical value of the aerodynamic Weber number, the vertical velocity of vapor $U_{cr}$ for the explosion can be calculated given that the length $d_o$ scales as $\sim 2R_c$, the contact area diameter just prior to the explosion. The mass of vapor generated through evaporation can be expressed as $\dot{m}_{gen}$ (Eqn. 6) and equated to the vapor flow immediately underneath the drop, $\rho_v A U_{cr}$ where $A(= \pi R_c^2)$ is the surface area of below the drop. This yields the following expression for the temperature required for the transition to the jet formation regime.

$$T_{c,II} \approx T_{sat} + \frac{\rho_v hL}{k_{eff}} \sqrt{\frac{We_{cr}\sigma_l}{\rho_v d_0}} \qquad (9)$$

A comparison between the theoretical temperature (Eqn. 9) and experimental temperatures at which the splashing to jet-splashing transition occurs shows a remarkable degree of match for different spacings (Fig. 4).

**From jet-splashing to Leidenfrost regime.** As the temperature of the substrate is increased in the jet-splashing regime, the thickness of the vapor cushion below the drop increases which relaxes the pressure and ultimately leads to a stage where the drop levitates completely (LFP). Unlike the transition for Curve II, the pressure required for the critical acceleration of the interface can be precisely calculated for the transition from jet-splashing to Leidenfrost regime.

Asymptotically, the pressure below the drop just before the LFP and immediately after it must be almost the same. This is equivalent to converging towards the LFP from higher temperature by decreasing the temperature and from lower temperature by gradually increasing the temperature.

Assuming the pressure to be $\Delta P$ and its corresponding surface area $A(=\pi R_c^2)$ before the jetting happens, the acceleration $a_{lv}$ in Eqn. (8) can be written as $\sim \Delta P \pi R_c^2/m$, where the total vapor mass ($m$) below the droplet before the jetting is given by $m = \dot{m}_{jet}\tau_{jet,exp}$ where $\tau_{jet,exp}$ is the time at which jetting begins. Substituting the acceleration from Eqn. (8) leads to the expression for the total jetting pressure as:

$$\Delta P_{jet} \approx \frac{12\pi \dot{m}_{jet}\tau_{jet,exp}\sigma_l}{R_c(b,\tau_{jet})^2 \rho_l \lambda_c^2} \qquad (10)$$

where, the $\lambda_c$ (~$4R_c$) is the wavelength of the RT instability and $R_c(b, \tau_{jet})$ is the droplet contact radius at the instant of jet formation for a given spacing. $\dot{m}_{jet}$ is calculated using the measured jet velocity at the very interface before the explosion[56,57]. Eqn. (10) provides an estimate of the pressure below the drop as we approach the LFP from the jet formation regime. Our next endeavor is to estimate the same pressure as we approach the LFP from higher temperatures. To do so, we consider Poiseuille flow at LFP[5,58], from which we obtain the pressure, $\Delta P_{LFP}$ to be $\sim 3\mu_v \dot{m}_{LFP}/2\pi\rho_v h^3$. We can use the same expression we used for $\dot{m}_{gen}$ for $\dot{m}_{LFP}$, using $k_v$ as the vapor conductivity which is the only medium in contact with the interface at the Leidenfrost state. $\mu_v$ represents the dynamic viscosity of the vapor at $T_{sat}$. Finally, equating the expressions for $\Delta P_{jet}$ and $\Delta P_{LFP}$, the temperature at which the drop shows no jetting and transitions to the Leidenfrost state can be found using the expression below:

$$T_{c,III} \approx T_{Sat} + \left(\frac{\pi \dot{m}\tau_{jet,exp}\sigma\rho_v h_{fg} h^4}{\mu k_v R_c^6(b,\tau_{jet})\rho_l}\right) \qquad (11)$$

The jet-splashing/LFP transition is captured accurately by Eqn. (11) and compares well with experimental data (Fig. 4).

**Discussion**

In conclusion, we investigated the boiling transitions on textured surfaces at micropillar spacing ranging from 1 to 100 μm, starting from the saturation temperature ($T_{sat}$) up until a few degrees above the Leidenfrost point ($T_{LFP}$). The stability of Leidenfrost vapor films in pool boiling are known to be dependent on surface wetting[59]. Our results indicate that the LFP curve for impacting drops is also a function of the post spacing, encountering a minimum at a specific spacing (~10μm in our case) between the micropillars. We report a previously unknown increase in the Leidenfrost temperature at lower spacings ($b<10$μm) and a surprising ejection of a vapor jet which pierces the droplet. To understand these observations, we carefully examined the underlying dynamics leading up to the Leidenfrost point and proposed a new mechanism based on hydrodynamic instabilities. We hypothesized the preponderance of these instabilities at smaller spacings ($b< 10$ μm) as opposed to boiling related mechanisms which explain droplet behavior at higher post spacings ($b>10$μm). We have studied the different boiling transitions using X-Ray phase contrast imaging which provides unprecedented visualization of the boiling morphologies, giving credence to our hypothesis that these transitions are indeed strongly influenced by hydrodynamic instabilities. To quantitatively predict these transitions, we developed mathematical models for each boiling regime and compared their predictions with the experimental observations and find them to be in good agreement with each other. Our findings also deepen understanding on the topic, which had posited enhanced cooling on substrates which exhibit increased Leidenfrost temperatures. It had been suggested that more surface area, as a consequence of more roughness, is responsible for less heat transfer needed to reach LFP in small spacing[20]. Although this is true for spacings above 10 μm and is explained in prior literature, extending the range of investigation to micropillar spacings below 10 μm, we see that higher LFPs are recorded for almost the same temperature for post spacings above and below 10 μm. Furthermore, in smaller spacings, more surface area to achieve a higher heat transfer rate (i.e. cooling), is provided. However, the instability-driven mechanisms (KH and RT), due to the significant pressure build up and high vapor velocities, decrease the contact between the droplet and substrate, while showing behaviors such as splashing and jet formation. Consequently, these instabilities may prevent the previously used force balance approach to accurately predict the $T_{LFP}$ (See supplementary Note 5). This leads to the conclusion that increased $T_{LFP}$ in small spacings does not necessarily guarantee a higher heat transfer required for cooling applications. Our novel findings provide a general framework needed in future studies. To begin with, determining the pressure build-up below the droplets on superheated textured surfaces may hold the key for understanding the different transitions. Secondly, while in our models we have relied on using the experimentally obtained droplet contact radius to calculate the different transitions, further studies are needed to accurately predict the evolution of contact radius on superheated textured surfaces. Our results are expected to impact a diverse range of applications thus underscoring their relevance. For example, we anticipate that our results would lead to better fire safety and containment measures, specifically those which involve spraying of water jets. Besides, our studies could guide optimal design of solid surfaces for cooling-related applications in power plants, nuclear reactors, and metallurgy-related applications involving quenching to obtain the alloy microstructures with desired mechanical properties such as toughness and hardness.

**Methods**

**Fabrication of micro-structured surfaces.** First, <100> 500μm thick SSP silicon wafers were cleaned with acetone, ethanol and isopropyl alcohol, rinsed with deionized water and dried with $N_2$ gas. The wafer is dehydrated under a hot plate at 115°C. Photoresist (MRS 703) is spin-coated onto the wafer at 3200 rpm. The wafer is exposed to a dose of 85 MJ/cm$^2$ and Ultraviolet light of wavelength 405 nm in the Heidelberg MLA 150 Direct Write Lithographer. After the Post process baking at 95°C, the pattern is developed with AZ 400 MIF and rinsed with water. The wafer is further subjected to the Plasmatherm Deep Silicon Etch to a recipe of $C_4F_8$ (150 sccm), $SF_6$ (150 sccm), Ar (30 sccm) for etching silicon microposts of 10 μm depth. The remaining photoresist was stripped off with NMP solution at 80°C for 15 minutes. Then, the wafer was rinsed with deionized water and dried off with $N_2$ gun. For the preparation of Nano-grass, a similar cleaning procedure was followed. The etch conditions in the Deep Silicon Etch were $SF_6$ (70 sccm) and $O_2$ (50 sccm). Finally, the 4-inch wafer was cut into 1-inch square substrates using the TYKMA Laser Marking system. Samples were plasma cleaned (Herrick plasma cleaner) providing a superhydrophilic surface for the drop deposition experiments.

**Surface Characterization.** The pillars height, diameter and the spacing between them were measured using Scanning Electron Microscopy and were confirmed to match the drawing for Direct Write Lithographer. The spacing *b* was considered to be the distance between outer walls of micropillars. Contact angle *θ* for the hydrophilic samples were measured to be around ~5-20º using a goniometer. This value is expected even after plasma treatment, considering the organic absorbing nature of Silicon.

**LFP Measurements.** Samples were firmly placed to a hot plate heater providing variable heat flux to control and maintain the temperature. The temperature of silicon substrates was measured and calibrated using thermocouples attached on top and underneath the wafers. Droplets ranging from 10 to 30μL were deposited using a droplet deposition system specifically designed for this experiment. The distance between the droplet and the substrate varied from ~0 (gentle touch) to 5mm resulting $We$ ~1 to 8. The experimental setup schematic is shown in Supplementary (Fig. S1a).

**XRPCI (X-Ray Phase Contrast Imaging).** Synchrotron X-Ray phase contrast imaging system with XSD 32-ID beam line was used to record videos at 20,000fps to observe the structures inside the droplet on the hot substrates. The field of view was 2×2 mm$^2$ with resolution of 2μm/pixel. Photron high speed camera was used to observe the splashing behavior of the droplet (Fig. S1c).

**Acknowledgements**

The fabrication of silicon surfaces was performed at Pritzker Nanofabrication Facility (University of Chicago). The SEM of samples was obtained at the Electron Microscopy Service (Research Resources Center, UIC). SA thanks the financial support of UIC College of Engineering. This research used resources of the Advanced Photon Source, a U.S. Department of Energy (DOE) Office of Science User Facility operated for the DOE Office of Science by Argonne National Laboratory under Contract No. DE-AC02-06CH11357.

**Author Contributions**

S.A. and N.P. conceived the idea of investigating boiling on nano/micro-structured surfaces. N.S., V.K. and S.A. designed the experiments. N.S. conducted the experiments in lab. V.K. and S.A. conducted experiments at ANL with help from K.F. V.K., N.S. and S.A. developed the theoretical models. All authors interpreted the results. N.S., V.K. and S.A wrote the manuscript and all authors provided the comments. N.S and V.K had equal contribution as first author to the work.

**Additional Information**

Supplementary information is available per request. Requests for supplementary materials should be addressed to N.S. (nsanei2@uic.edu) or the corresponding author S.A. (sushant@uic.edu).

**Competing Interests**

The authors declare that they have no competing interests.